\documentclass[nature,onecolumn,superscriptaddress,floatfix]{revtex4-1}

\usepackage{epsfig,dcolumn,amsmath,latexsym}
\usepackage{graphicx}
\usepackage{subfigure}
\usepackage[normalem]{ulem}
\usepackage{cancel}
\usepackage{multirow}

\usepackage{color, soul}

\begin{document}

\title{Electronic structure of Ni-doped EuRbFe$_4$As$_4$: Unique crystal field splitting and multiband RKKY interactions}

\author{Chenchao Xu}
 \affiliation{Department of Physics, Zhejiang University, Hangzhou 310027, China}

\author{Qijin Chen}
 \email[E-mail address:] {qjchen@zju.edu.cn}
 \affiliation{Department of Physics, Zhejiang University, Hangzhou 310027, China}

\author{Chao Cao}
 \email[E-mail address: ]{ccao@hznu.edu.cn}
 \affiliation{Condensed Matter Group,
  Department of Physics, Hangzhou Normal University, Hangzhou 311121, China}

\date{\today}
\begin{abstract}
The relationship between magnetism and superconductivity has been one of the most discussed topics in iron-based superconductors. Using the first-principles calculations, we have studied the electronic structure of 1144-type iron-based superconductor EuRbFe$_4$As$_4$. We find the crystal field splitting of EuRbFe$_4$As$_4$ is unique, such that the $d_{z^2}$ orbitals are closer to the Fermi level $\epsilon_F$ than the $d_{xy}$ orbitals. The RKKY interaction strength is estimated to be approximately 0.12 meV in prestine EuRbFe$_4$As$_4$. Upon Ni-doping on the Fe site, the RKKY interaction strength is barely changed upon Ni doping due to the highly anisotropic FSs and multiband effect, despite of the drastically reduced $d_{zx(y)}$ DOS at $\epsilon_F$. Finally, in both pristine and doped compounds, the RKKY interaction is primarily mediated through bands due to Fe-$d_{z^2}$ orbitals.Our calculations suggest the RKKY interaction mediated by $d_{z^2}$ orbital is probably responsible for the magnetism in EuRbFe$_4$As$_4$ and doesn't change upon Ni-doping.
\end{abstract}

\maketitle

\thispagestyle{empty}

\section*{Introduction}

Over the last decade, enormous efforts have been put on the iron-based superconductors (FeSCs) since the first discovery of superconductivity in F doped LaOFeAs in 2008\cite{FeSC:LaOFFeAs_26K_Kamihara}. To date the relationship between magnetism and superconductivity has been extensively investigated both experimentally and theoretically to pursue the essence of pairing mechanism. As a matter of fact, a multitude of the FeSCs are magnetic metals with quasi-2D electron and hole pockets. Usually the parent compounds of FeSCs are non-superconducting and undergo structural transition from tetragonal lattice to orthorhombic lattice followed by a spin density wave (SDW) transition when the temperature is decreased. Once the system is doped, however, both structural and SDW transitions are suppressed and superconductivity emerges\cite{0034-4885-74-12-124508,RevModPhys.83.1589}.

Recently a new type of FeSC structure AeAFe$_4$As$_4$ (AeA-1144; Ae=Ca,Sr; A=K,Rb,Cs) has been successfully synthesized by A. Iyo et al. \cite{FeSC:AeAFeAs_30K_Iyo} with superconducting transition temperature $T_c>$30K without doping. Despite of the same stoichiometric ratio, these 1144-type compounds are different from the 50\% K-doped BaFe$_2$As$_2$ compounds such that the Ae atoms are separated from the A atoms in different layers, and therefore the structure can be viewed as layer-by-layer stacked AeFe$_2$As$_2$ (Ae-122) and AFe$_2$As$_2$ (A-122). Independly, Liu et al. and Kenji et al. \cite{FeSC:RbEuFeAs_36K_YiLiu,JPSJ.85.064710} then managed to make an intergrowth structure of RbFe$_2$As$_2$ and EuFe$_2$As$_2$ leading to another 1144-type compound EuRbFe$_4$As$_4$. Similar to the AeAFe$_4$As$_4$ (Ae=Ca,Sr;A=K,Rb,Cs) compounds, EuRbFe$_4$As$_4$ exhibits superconductivity at 35K without any doping. In addition, ferromagnetic behavior due to the Eu-layers was observed below 15K without destroying the  superconductivity, suggesting a robust coexistence of superconductivity and ferromagnetism \cite{FeSC:RbEuFeAs_36K_YiLiu}. The 4$f$ electrons are nearly fully localized, yielding a large moment of 6.5 $\mu_B$/Eu. Interestingly, the Curie temperature was robust against the Ni-doping, while the superconducting $T_{sc}$ was very sensitive\cite{PhysRevB.96.224510}. It was therefore conjectured that the $d$-$f$ superexchange or Eu-As-Eu superexchange is likely to be the mechanism mediating the magnetic interactions.

In this article, we present a systematic first-principles study of EuRbFe$_4$As$_4$ compound. We compare EuRbFe$_4$As$_4$ with the CaRbFe$_4$As$_4$ as well as the traditional BaFe$_2$As$_2$ compound. The band structure of EuRbFe$_4$As$_4$ resembles that of CaRbFe$_4$As$_4$, and there are 10 bands crossing the Fermi level in both compounds. However, for EuRbFe$_4$As$_4$, the $d_{z^2}$-band, which is much lower than the Fermi level in other FeSCs, is now elevated close to $\epsilon_F$ around $\Gamma$ point, thus the undoped EuRbFe$_4$As$_4$ system is very close to a Lifshitz-transition. As a result, the $d_{z^2}$-orbital is more relevant in the process of electrical transport and magnetism in the EuRbFe$_4$As$_4$ compound. By decomposing the $J_{\mathrm{RKKY}}$ into contributions from different Fermi surface sheets, we find this interaction is primarily mediated by the outmost hole pocket due to $d_{z^{2}}$ orbital around the $\Gamma$ point, suggesting that this orbital is important to the Eu-magnetism. Under nickel doping, the RKKY exchange interaction is little affected, despite of the drastically reduced density of state from $d_{xz}$ and $d_{yz}$ orbitals. This may serve as a possible explanation why the magnetic transition $T_{FM}$ is relatively constant while the superconductivity $T_{sc}$ is quickly suppressed by Ni-doping in this compound. 

\section*{Results and Discussion}
\subsection{Crystal Structure and Electronic Structure}

\begin{figure}[h]
  \includegraphics[width=15 cm]{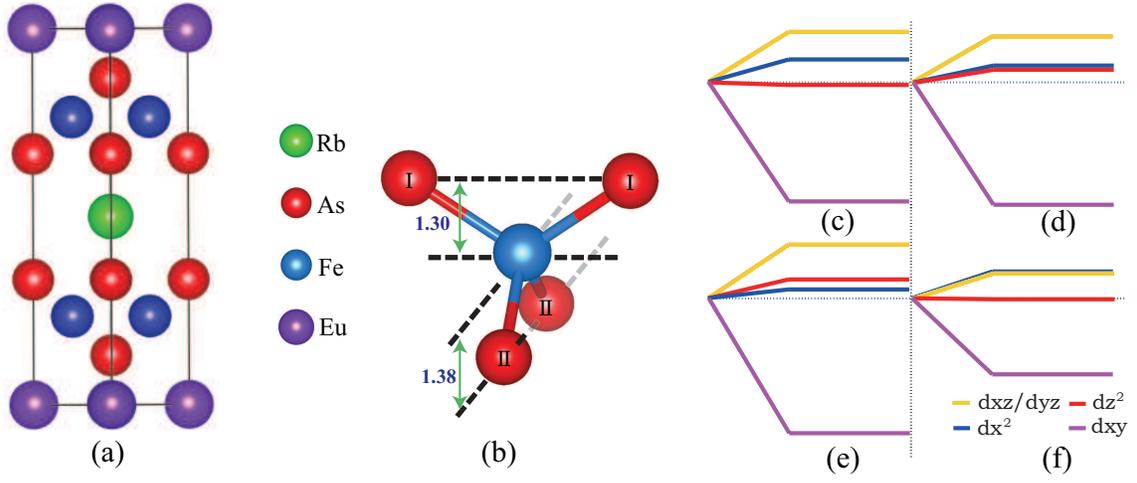}
  \caption{Crystal structure and crystal field splitting. (a) Crystal structure of EuRbFe$_4$As$_4$. (b) FeAs layer centred on Fe atom. Compared with 122-type EuFe$_2$As$_2$, the Rb-substitution breaks not only the (1/2 1/2 1/2) gliding symmetry but also the S$_{4}$ symmetry. (c-f) Crystal field splitting of Fe $3d$-orbitals in (c) BaFe$_2$As$_2$ (PM), (d) CaRbFe$_4$As$_4$ (PM), (e) EuRbFe$_4$As$_4$ (FM) and (f) EuRbNi$_4$As$_4$ (FM) respectively. \label{fig:CS_CEF}}
\end{figure}

The crystal structure of EuRbFe$_4$As$_4$ is tetragonal P4/mmm with lattice parameters $a$=3.89\AA\ and $c$=13.31\AA\ as illustrated in Fig. \ref{fig:CS_CEF}. The crystal consists of FeAs-layers separated by alternating Rb-layers and Eu-layers. Therefore each unit cell contains 2 FeAs layers that are mirror-symmetric with respect to either the Rb-layer or the Eu-layer. In contrast to most traditional FeSCs, the two intercalating layers sandwiching each FeAs-layer are different in 1144-type FeSCs, thus the Fe atom is no longer located at the center of the tetrahedral formed by the 4 closest As atoms (Fig. \ref{fig:CS_CEF} a-b). Therefore, the S$_4$ local symmetry around each Fe is reduced to C$_2$ in 1144-type FeSCs.

\begin{figure}[h]
  \includegraphics[width=15 cm]{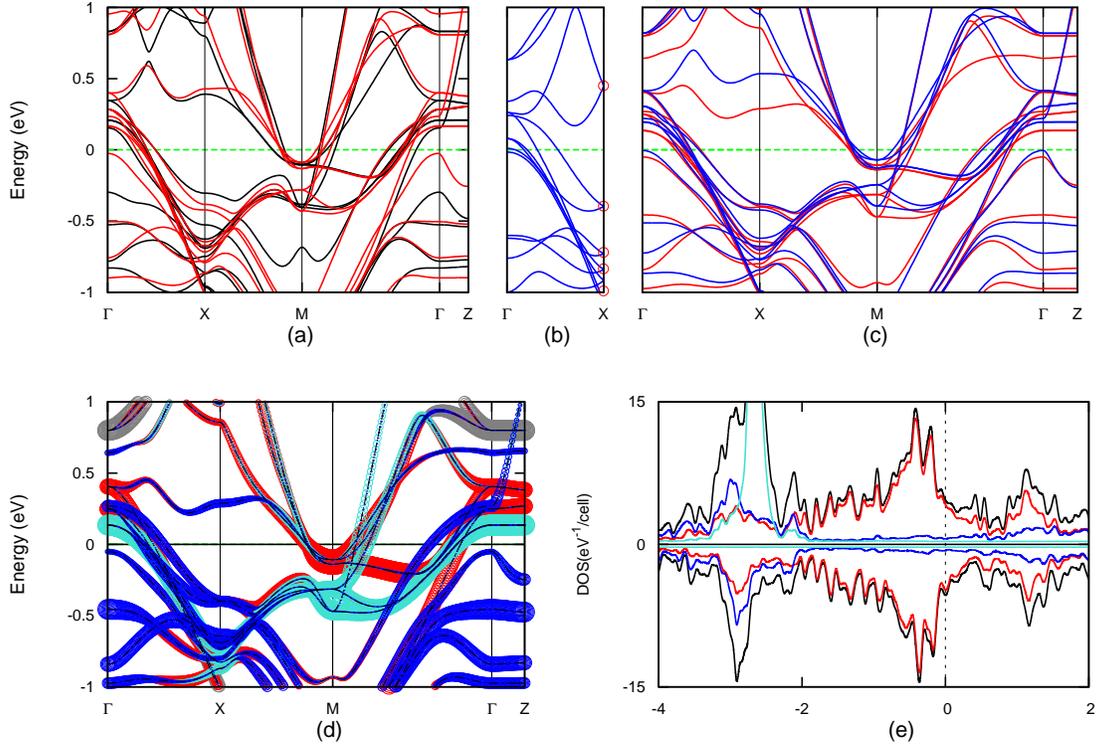}
  \caption{Electronic band structure and density of states. (a) Electronic band structure of EuRbFe$_4$As$_4$(red line) compared with CaRbFe$_4$As$_4$(black line). (b) Electronic band structure of BaFe$_2$As$_2$(blue line) in PM state in the folded Brillouine zone. The green dotted line denotes the Fermi level, which is aligned at zero. The degenered points at X point as indicated by red circles in BaFe$_2$As$_2$ are removed in 1144-system.  (c) Spin resolved band structure of EuRbFe$_4$As$_4$ in FM state. The red and blue lines correspond to the two spin channels respectively. (d) Projected Band structure of EuRbFe$_4$As$_4$ (FM phase) for majority spin channel around Fermi level. The size of the red, turquoise, blue and gray circles are proportional to the contributions from the $d_{x(y)z}$, $d_{xy}$, $d_{z^{2}}$ and $d_{x^{2}-y^{2}}$ respectively. (e) Total electronic density of state and projected density of state of EuRbFe$_4$As$_4$.The black, red, blue and turquoise lines represent the total DOS, the PDOS for Fe-$d$, As-$p$, and Eu-$f$ orbitals respectively. The Fermi level is aligned at zero.\label{fig:bands_DOS}}
\end{figure}

\begin{figure}[h]
  \includegraphics[width=15 cm]{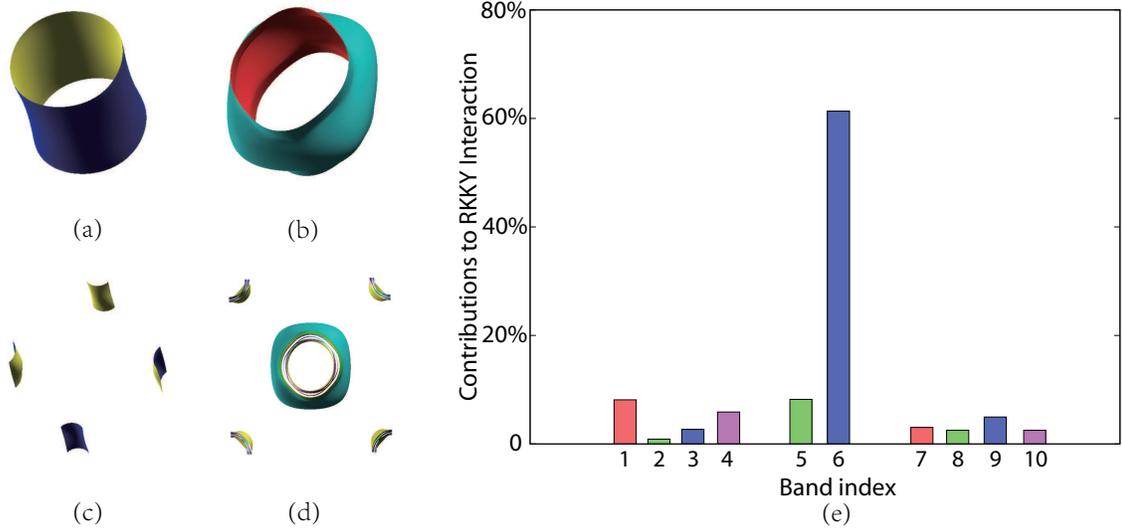}
  \caption{Fermi surfaces and their relative contributions to RKKY interactions. (a-d) Fermi surfaces of EuRbFe$_4$As$_4$. (a) typical hole pocket due to $d_{xy}$ with little dispersion along $k_z$ (b) typical hole pockets due to $d_{z^2}$ with strong dispersion along $k_z$; (c) typical electron pockets due to $d_{zx}$/$d_{zy}$; and (d) all pockets combined together. The zone center is $\Gamma$ in all panels. (e) Band-decomposed contribution to RKKY interaction in the pristine EuRbFe$_4$As$_4$. The band indices are labeled from lowest to highest at the $\Gamma$ point. There are 10 bands crossing the Fermi levels, among which 6 are hole-like and 4 are electron-like. The 6$^{th}$ band with the largest contributions corresponds to the $\Gamma$-centred outmost hole pocket due to $d_{z^2}$ shown in (b).  \label{fig:fs_bands_contributions}}
\end{figure}

The band structures of EuRbFe$_4$As$_4$ (red line), CaRbFe$_4$As$_4$ (black line) and folded BaFe$_2$As$_2$ (blue line, $\Gamma$-X) in paramagnetic (PM) phase are shown in Fig.  \ref{fig:bands_DOS}a-b. Both EuRbFe$_4$As$_4$ and CaRbFe$_4$As$_4$ exhibit the same number of hole pockets around $\Gamma$ points and electronic pockets around M points, and the dispersions resemble each other. Nevertheless, the highest occupied state at $\Gamma$ is extremely close to the Fermi level $\epsilon_F$ in EuRbFe$_4$As$_4$ (while in CaRbFe$_4$As$_4$ it is more than 200 meV below $\epsilon_F$), suggesting the undoped EuRbFe$_4$As$_4$ system is on the edge to a Lifshitz-transition. Projected band structure shows that the orbital character of this state is $d_{z^{2}}$ (Fig. \ref{fig:bands_DOS}d). With slight hole-doping (0.5 hole per f.u.), this band will cross Fermi level and create a new Fermi surface sheet. In Ba$_{1-x}$Rb$_{x}$Fe$_{2}$As$_{2}$, it was reported that a crossover from nodeless to nodal SC occurs at $x$=0.65 as a result of a hole-doping induced Lifshitz transition\cite{PhysRevB.93.094513}. Considering the similarities between these systems, such transition may also be present in hole-doped EuRbFe$_4$As$_4$. Additionally, in contrast to BaFe$_2$As$_2$, owning to the S$_{4}$ symmetry breaking, the bands structures in 1144-system splits along X-M. Similar splitting is also present in CaKFe$_4$As$_4$ \cite{CaKFeAs_Lochner}. We have also calculated EuRbFe$_4$As$_4$ (Fig. \ref{fig:bands_DOS}c) in the non-superconducting FM phase, where Eu moments align in parallel. The exchange spin splitting is less than 20 meV shown in Table \ref{tab:splitting}. The projected band structure (Fig. \ref{fig:bands_DOS}d) indicates that the $d_{z^{2}}$ orbital has sizable contribution in EuRbFe$_4$As$_4$, in addition to the $d_{xz}$, $d_{yz}$ and $d_{xy}$ orbitals that universally dominate the electronic states near $\epsilon_F$ in FeSCs \cite{PhysRevB.81.014526,PhysRevB.83.054510}. The $d_{x^2-y^2}$ orbital strongly hybridizes with As-$4p$, and is present beyond $\epsilon_F\pm1$ eV range. Six bands cross the Fermi level around the $\Gamma$ point, forming the hole pockets around the Brillouin zone (BZ) center (Fig. \ref{fig:fs_bands_contributions} a-b). In addition, another 4 bands cross the Fermi level around the M($\pi $,$\pi $) point forming the electron pockets corresponding to Fermi surfaces (FSs) in Fig. \ref{fig:fs_bands_contributions}c.

The density of state (DOS) of EuRbFe$_4$As$_4$ in FM phase are illustrated in Fig. \ref{fig:bands_DOS}e. Similar to all FeSCs, the DOS of FeSCs near $\epsilon_F$ is dominated by Fe-$3d$ orbitals, hybridized with As-$4p$ orbitals.  The total DOS at $\epsilon_F$ is 9.41 eV$^{-1}$/f.u., equivalent to electronic specific-heat coefficient $\gamma$ = 21.642 mJ/(K$^{2}\cdot$ mol). Noticing that $g(\epsilon_F)$ of CaRbFe$_4$As$_4$ is 9.78 eV$^{-1}$/f.u. \cite{DFT_CaRbFeAs}, similar to what we obtained for EuRbFe$_4$As$_4$. As a comparison, the experimental value of $g(\epsilon_F)$ is 60 eV$^{-1}$/f.u. \cite{FeSC:RbEuFeAs_36K_YiLiu}, indicating a large electron mass renormalization factor of $\approx$6.4. From spin resolved DOS (Fig. \ref{fig:bands_DOS}e) the majority-spin channel/minority-spin channel of $4f$-electrons are located around 2.7eV(10eV) below(above) $\epsilon_F$. By integrating the $4f$ contribution below $\epsilon_F$, each Eu atom has a local moment of 7$\mu_{B}$, consistent with the fully-localized nature of 4$f$ electrons in EuRbFe$_4$As$_4$.


\subsection{Single Particle Hamiltonian and Crystal Field}
To analyze the crystal field effect in EuRbFe$_4$As$_4$, we have employed the maximally projected Wannier function method \cite{PhysRevB.74.195118} to fit the DFT band structure. In order to capture the main features of the electronic structures of EuRbFe$_4$As$_4$, we take into account not only Fe-$3d$ orbitals and As-$4p$ orbitals, but Eu-$4d$ and Eu-$4f$ orbitals as well. Using these 44 orbitals not only allows us to describe the low energy physics of this system, but also elucidates the electronic hopping process between $d$-$p$ as well as $d$-$f$ orbitals, which are relevant in the superconducting and magnetic properties of this compound.

In Fig. \ref{fig:CS_CEF} (c-f) we compare the crystal field splitting between BaFe$_2$As$_2$, CaRbFe$_4$As$_4$, and undoped EuRbFe$_4$As$_4$ and the corresponding values are listed in Table \ref{tab:splitting}. In general, the electronic bands of $3d$-systems strongly depend on the crystal symmetry. Under perfect tetrahedral crystal field the 5 $d$-orbitals split into two groups: triply degenerate $t_{2g}$ levels and doubly degenerate $e_{g}$ levels. In the family of iron-based superconductors, all the members share the similar Fe-As layer with tetrahedral symmetry. Due to the combined effect of anion crystal fields and surrounding cations, the hybridization between the Fe-$d$ orbitals and the pnictogen-/chalcogen-$p$ orbitals appears. Therefore both $t_{2g}$ and $e_g$ orbitals further split, leaving only degenerate $d_{xz}$ and $d_{yz}$, as exemplified in BaFe$_2$As$_2$ (Fig. \ref{fig:CS_CEF}c).  In addition, we find in EuRbFe$_4$As$_4$ the energy levels of $d_{xy}$ and $d_{z^{2}}$ are reversed, consistent with the observation in the band structure results. Furthermore, the exchange spin splitting in EuRbFe$_4$As$_4$ is one order of magnitude smaller compared to the crystal field splitting, and the formation of long-range magnetic order does not affect magnitude and order of crystal splitting.

In Table \ref{tab:hopping} we show the important hopping terms related with low energy excitation for BaFe$_2$As$_2$,  CaRbFe$_4$As$_4$, and EuRbFe$_4$As$_4$ (both PM state and FM state) respectively. In BaFe$_2$As$_2$ compound, the As atoms above (As$^{\mathrm{I}}$) and below (As$^{\mathrm{II}}$) the Fe-plane are equivalent (S$_4$ local symmetry), thus the hoppings between Fe-$3d$ and either As-$4p$ orbitals are the same. However, this symmetry is broken in 1144-type compounds, thus the hoppings between Fe-$3d$ and As$^{\mathrm{I}}$-$4p$ and those between Fe-$3d$ and As$^{\mathrm{II}}$-$4p$ are no longer identical. For all these compounds the strongest hopping comes from Fe-3$d_{x(y)z}$ and As-4$p_{y(x)}$, followed by the hopping between Fe-3$d_{xy}$ and As-4$p_{z}$. Compare to the $d$-$p$ hopping, the $d$-$f$ hopping terms are one order smaller, among which the Fe-3$d_{z^{2}}$ and Eu-4$f_{xyz}$ is largest. Remarkably, the $d$-$p$ hopping terms in PM phase and FM phase are roughly the same for EuRbFe$_4$As$_4$, despite of the large magnetic moment of Eu atoms. Therefore, the effect of ferromagnetic Eu layers can be equivalently regarded as magnetic exchange splitting field for the FeAs-layers. It is worthy noting that the H$_{c2}$(0) of a similar compound CaKFe$_4$As$_4$ can reach as high as 71 T, \cite{PhysRevB.94.064501} while the internal field due to Eu$^{2+}$ atom was much smaller\cite{0953-8984-23-6-065701}. Therefore, the superconductivity may coexist with the FM Eu-layer.

\subsection{RKKY Exchange Interaction}
The Ruderman-Kittel-Kasuya-Yosida (RKKY) exchange interaction was reported to be inseparably related with the magnetic order in 122-type EuFe$_2$(As,P)$_2$ \cite{PhysRevB.79.155103,PhysRevB.82.094426} and 1144-type EuRbFe$_4$As$_4$ \cite{FeSC:RbEuFeAs_36K_YiLiu}. Since In the weak coupling limit, the RKKY interaction can be approximatively by $J_{\mathrm{RKKY}}\sim J_{\mathrm{K}}^2 g(\epsilon_F)$ \cite{DONIACH1977231,RKKY_CeFeAsPO,RKKY_EuFeAsP} with a bare Kondo coupling $J_{\mathrm{K}}$ in the ordered phase and density of states at the Fermi level $g(\epsilon_F)$. Following the Schrieffer-Wolff transformation \cite{SchriefferWolff,PhysRevB.80.020505}, and assuming large $U$ limit, we calculate the Kondo coupling strength by $$J_{0n}=\frac{1}{g_n(\epsilon_F)}\int d\mathbf{k} \frac{V_{n\mathbf{k},f}^*V_{f,n\mathbf{k}}}{|\epsilon_f-\epsilon_{n\mathbf{k}}|}\times\delta\left(\epsilon_{n\mathbf{k}}-\epsilon_F\right)$$ for the conduction channel $n$, where $V_{n\mathbf{k},f}=\langle w_f|\hat{H}|n\mathbf{k}\rangle=\sum\limits_{\alpha} \langle w_f|\hat{H}|w_{\alpha}\rangle \langle w_{\alpha}|n\mathbf{k}\rangle$. In these expressions, $n$ is the conduction band index, $|n\mathbf{k}\rangle$ is the $n$-th Bloch state of conduction electrons at $\mathbf{k}$, $|w_{\alpha}\rangle$ is the $\alpha$-th local wannier state for the conduction electrons, $\epsilon_{n\mathbf{k}}$ and $\epsilon_f$ are the eigen-energies of the conduction channel and local $f$ states, $g_n(\epsilon_F)$ is the density of states at the Fermi level of the $n$-th conduction channel, and $|w_f\rangle$ denotes the local wannier state of the $f$-electrons. The averaged Kondo coupling $\bar{J_K}$ is roughly -0.73 meV whose magnitude is slightly smaller than EuFe$_{2}$As$_{2}$ (-1.01 meV)\cite{RKKY_EuFeAsP}. Considering the multiband effect and highly anisotropic Fermi surfaces, the RKKY interaction is obtained by summing up the contributions from all conduction bands $J_{\mathrm{RKKY}}=\sum\limits_{n} J_{0n}^2 g_n(\epsilon_F)$. The resulting $J_{\mathrm{RKKY}}$ is around 0.12 meV corresponding to $T_{FM} \approx $12.5 K, which is in good agreement with the experimental value\cite{FeSC:RbEuFeAs_36K_YiLiu}. It's worth noting that the $J_{\mathrm{RKKY}}$ in our calculation is averaged in all directions. To tell the difference between the inter-layer coupling  $J_{\perp}$ and intra-layer coupling $J_{\parallel}$,  it is necessary to employ $J_{\mathrm{RKKY}}(\mathbf{r})=J_{K}^2\sum\limits_{\mu\nu}\chi^{\mu\nu}(\mathbf{r})$ (where $\chi^{\mu\nu}(\mathbf{r})=\sum\limits_{\mathbf{k},\mathbf{k'}}e^{i(\mathbf{k}-\mathbf{k'})\mathbf{r}}[\frac{f_{\mu}(\mathbf{k})-f_{\nu}(\mathbf{k'})}{\varepsilon_{\mathbf{k}}^{\mu}-\varepsilon_{\mathbf{k'}}^{\nu}}]$)\cite{PhysRevB.84.134513}. Using the Wannier orbital based tight-binding Hamiltonian, we have estimated $J_{\perp}/J_{\parallel}\approx1.0$. Such nearly isotropic behavior can be understood by calculating the band-decomposed contributions from each Fermi surface sheet (Fig. \ref{fig:fs_bands_contributions} e), which suggests more than 60\% of $J_{\mathrm{RKKY}}$ is mediated by the outmost hole pocket around $\Gamma$ point with prominent 3D feature. This Fermi surface sheet is mainly due to the Fe-$d_{z^{2}}$ orbitals, and a similar mechanism was also previously proposed in EuFe$_2$As$_2$ \cite{EuFeAs_ZhiRen}. 

It should be emphasized that the RKKY interaction in our calculation starts from the normal state. In superconducting state, the interplay between RKKY interaction and superconductivity is more complicated. It has been illustrated by N. Yao et al. that the antiferromagnetic contribution from indirect spin exchange will be enhanced due to the formation of Yu-Shiba-Rusinov (YSR) bound states, provided the binding energy of YSR state is close to the middle of superconducting gap $\Delta$\cite{PhysRevLett.113.087202}. In EuRbFe$_4$As$_4$, however, this binding energy is estimated to be close to $\Delta$, implying the weak hybridization between YSR state and superconducting condensate. Thus, the conventional RKKY interaction will still dominate in superconducting state.

\subsection{Effect of Nickel doping}
\begin{figure}[h]
  \includegraphics[width=15 cm]{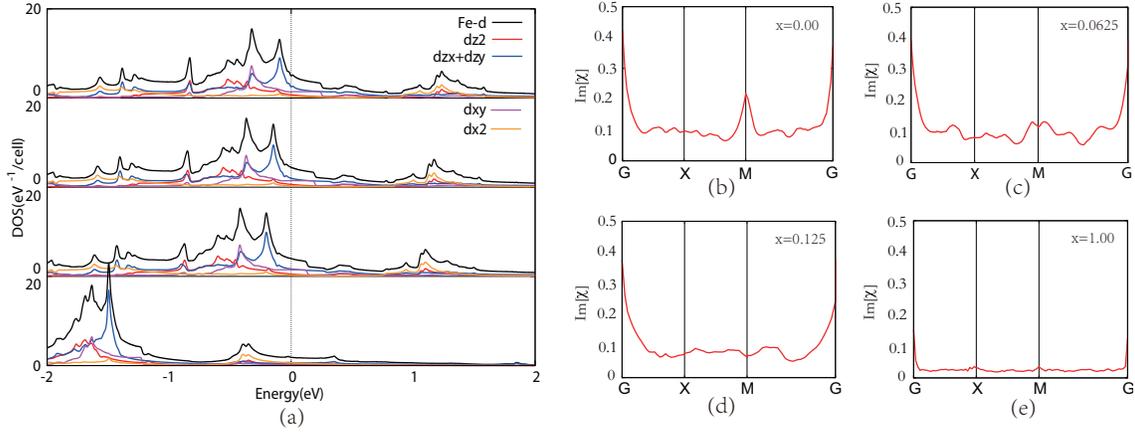}
  \caption{Effect of Ni-doping. (a) Total and projected density of states of EuRb[Fe$_{1-x}$Ni$_x$As]$_4$. The panels from top to bottom are undoped ($x$=0), $x$=0.0625, $x$=0.125, and $x$=1 (EuRbNi$_4$As$_4$), respectively. (b-e) Imaginary part of bare electron susceptibilities (nesting function). From (b) to (e) are undoped, 6.25\%-doped, 12.5\%-doped and fully doped EuRbFe$_4$As$_4$ respectively. \label{fig:doping_dos_wannchi}}
\end{figure}

In order to investigate the Ni-doping effects, we have also calculated EuRbFe$_{1-x}$Ni$_{x}$As$_4$ with $x$=6.25\%, 12.5\% and 100\% under the Virtual Crystal Approximation (VCA). The DOS of these system are illustrated in Fig. \ref{fig:doping_dos_wannchi} (a). With Nickel doping, the Fermi level moves to higher energy, demonstrating the main effect induced by Nickel is electron-doping. Comparing 6.25\% Ni-doping with the undoped EuRbFe$_4$As$_4$, the DOS of $d_{zx(y)}$, $d_{z^{2}}$ and $d_{x^{2}-y^{2}}$ orbitals are suppressed by 17.6\%, 13.7\% and 12.1\% respectively while the DOS of $d_{xy}$ is enhanced by 4.0\%. Note that the $T_{sc}$ for the superconductivity is usually very sensitive to the DOS at Fermi level $g(\epsilon_F)$. Therefore, the large decreased $g(\epsilon_F)$ of $d_{zx(y)}$ is consistent with strongly suppressed superconducting temperature upon doping. 

We have also calculated the imaginary part of bare electron susceptibility (nesting function) under different doping levels (Fig. \ref{fig:doping_dos_wannchi} (b-e)). For pristine EuRbFe$_4$As$_4$, the nesting function shows a sharp peak at M ($\pi$, 0), which signals large spin-fluctuation that considered to be related with the superconductivity\cite{PhysRevLett.101.057003,doi:10.1143/JPSJ.77.114709,PhysRevB.82.180501}. This peak is quickly suppressed by Ni-doping, and is already absent with 12.5\% Ni-doping, suggesting that the superconductivity can be quickly suppressed by Ni-doping. The imaginary part of the 100\%-doped compound (or EuRbNi$_4$As$_4$) is completely flat, meaning that the Ni-lattice is without spin-fluctuation. In fact, our DFT calculation confirms that Ni-lattice is completely nonmagnetic in EuRbNi$_4$As$_4$.

To evaluate the influence of doping on the crystal field splitting and the RKKY interaction, we naively calculate the 100\% Ni-doping situation. With all the irons substituted, the energy level of $d_{xy}$ orbital rises, substantially higher than $d_{z^{2}}$ orbital and even nearly degenerate with $d_{zx(y)}$ orbitals (Fig. \ref{fig:CS_CEF} f). Considering the main effect of Ni-doping at small $x$ is to shift the Fermi level upward, the averaged Kondo coupling $\bar{J_K}$ and the RKKY interaction $J_{\mathrm{RKKY}}$ are also calculated (Table \ref{tab:RKKY}) with the rigid band model (RBM). The averaged Kondo coupling $\bar{J_K}$ slightly decreases upon doping within RBM while the variation in VCA is nonmonotonic. Both of these two methods, however, show the RKKY interaction is around 0.1 meV, insensitive to the Ni-doping, which is in agreement with experimental observations. \cite{PhysRevB.96.224510}. 

In conclusion, we have performed first principles calculation on a typical 1144-type iron pnictide EuRbFe$_4$As$_4$. We analyze the detailed electronic band structure, density of state and Fermi surface as well as the distinction of electronic properties between EuRbFe$_4$As$_4$, BaFe$_2$As$_2$ and CaRbFe$_4$As$_4$. Through detailed analysis of the crystal splitting with tight-binding Hamiltonian, the energy levels of $d_{z^{2}}$ and $d_{xy}$ are reversed in EuRbFe$_4$As$_4$ which makes the $d_{z^{2}}$ orbital more important in electrical and magnetic transport. In addition, an extra electron pocket around the $\Gamma$ point is mainly from this orbital, which is close to a Lifshitz-transition. Upon Ni-doping, the DOS of $d_{zx}$/$d_{zy}$ orbitals at Fermi level as well as the spin fluctuation at ($\pi$,0) will be substantially suppressed, and both effects are detrimental to superconductivity. Finally, the RKKY interaction between Eu-layers is little affected by Ni-doping, and is mostly mediated through the outmost hole FS pocket around the $\Gamma$ point due to $d_{z^{2}}$ orbital.

\section*{Methods}
The calculations were carried out based on density functional theory (DFT) with Quantum ESPRESSO (QE) \cite{QE,PhysRevB.54.11169,PhysRevB.59.1758}. Throughout the calculations, the Perdew, Burke, and Ernzerhof parametrization (PBE) of generalized gradient approximation (GGA) to the exchange correlation functional was used\cite{GGA}. The energy cutoff of plane-wave basis was chosen to be 122 Ry (1220 Ry for the augmentation charge), which was sufficient to converge the total energy to 1 meV/atom. A $\Gamma$-centred 12$\times$12$\times$3 Monkhorst-Pack \cite{Monkhorst-Pack} k-point mesh was chosen to sample the Brillouin zone in the calculations. For the calculations of paramagnetic phases, the Eu-4$f$ states were considered to be fully localized core states and do not hybidize with any conduction electrons; while for the ferromagnetic state calculations, the Eu-4$f$ states were explicitly considered as semi-core valence states, with an additional Hubbard-like effective interaction $U=$7 eV \cite{LDAU,DFTU,LSDAU}. In addition, for ferromagnetic state calculations, only the initial magnetic moments on Eu atoms were set to non-zero values.

We have employed the experimental crystal structure to perform all the calculations, since a full structural relaxation will yield a 5.1\% shorter $c$ and wrong As-heights. This problem was known in other FeSCs\cite{PhysRevLett.100.237003,Ma2010}, and was due to the fact that spin fluctuations cannot be accounted for in the static mean-field implementation of DFT\cite{PhysRevB.78.085104}.

The DFT results were then fitted to a tight-binding (TB) model Hamiltonian with maximally projected Wannier function method\cite{Wannier90_code,PhysRevB.74.195118}. Although TB Hamiltonian constructed with 5 Fe-$3d$ orbitals is sufficient to describe the low-energy excitations in FeSCs, one must include Eu-$4f$ orbitals and other relevant ones in EuRbFe$_4$As$_4$ in order to analyze the crystal field splitting and magnetic exchange coupling interactions. Therefore, we have employed 44 atomic orbitals including the Fe-$3d$, As-$4p$, Eu-$5d$ and Eu-$4f$ orbitals in the fitting procedure.

\section*{Data availability}
The data that support the findings of this study are available from the corresponding author upon reasonable request.

\section*{Acknowledgements}

  The authors would like to thank Guanghan Cao, Jianhui Dai, Zhuan Xu, Yi Zhou and Fanlong Ning for the discussions and their valuable comments and suggestions. This work has been supported by NSFC (Grant Nos. 11274006, 11274267, 11774309 and 11874137) and 973 Project of MOST (Grant No. 2014CB648400). All calculations were performed on Tianhe-2 Superconducting Center of China as well as the High Performance  Computing Center of College of Science at Hangzhou Normal University.

\section*{Author contributions statement}

C.C.X. performed most of the calculations; C.C.X. and C. C. were responsible for the data analysis and drafted the manuscript; all authors participated in discussions.

\section*{Additional information}

\textbf{Competing financial interests} The authors declare no competing financial interests.

\bibliographystyle{naturemag}

\begin{thebibliography}{}
\expandafter\ifx\csname url\endcsname\relax
  \def\url#1{\texttt{#1}}\fi
\expandafter\ifx\csname urlprefix\endcsname\relax\def\urlprefix{URL }\fi
\providecommand{\bibinfo}[2]{#2}
\providecommand{\eprint}[2][]{\url{#2}}

\end{thebibliography}


\begin{thebibliography}{10}
\expandafter\ifx\csname url\endcsname\relax
  \def\url#1{\texttt{#1}}\fi
\expandafter\ifx\csname urlprefix\endcsname\relax\def\urlprefix{URL }\fi
\providecommand{\bibinfo}[2]{#2}
\providecommand{\eprint}[2][]{\url{#2}}

\bibitem{FeSC:LaOFFeAs_26K_Kamihara}
\bibinfo{author}{Kamihara, Y.}, \bibinfo{author}{Watanabe, T.},
  \bibinfo{author}{Hirano, M.} \& \bibinfo{author}{Hosono, H.}
\newblock \emph{\bibinfo{journal}{Journal of the American Chemical Society}}
  \textbf{\bibinfo{volume}{130}}, \bibinfo{pages}{3296--3297}
  (\bibinfo{year}{2008}).


\bibitem{0034-4885-74-12-124508}
\bibinfo{author}{Hirschfeld, P.~J.}, \bibinfo{author}{Korshunov, M.~M.} \&
  \bibinfo{author}{Mazin, I.~I.}
\newblock \bibinfo{title}{Gap symmetry and structure of Fe-based
  superconductors}.
\newblock \emph{\bibinfo{journal}{Reports on Progress in Physics}}
  \textbf{\bibinfo{volume}{74}}, \bibinfo{pages}{124508}
  (\bibinfo{year}{2011}).

\bibitem{RevModPhys.83.1589}
\bibinfo{author}{Stewart, G.~R.}
\newblock \bibinfo{title}{Superconductivity in iron compounds}.
\newblock \emph{\bibinfo{journal}{Rev. Mod. Phys.}}
  \textbf{\bibinfo{volume}{83}}, \bibinfo{pages}{1589--1652}
  (\bibinfo{year}{2011}).

\bibitem{FeSC:AeAFeAs_30K_Iyo}
\bibinfo{author}{Iyo, A.} \emph{et~al.}
\newblock \bibinfo{title}{New-structure-type fe-based superconductors:
  CaAFe$_4$As$_4$ (A = K, Rb, Cs) and SrAFe$_4$As$_4$ (A = Rb, Cs)}.
\newblock \emph{\bibinfo{journal}{Journal of the American Chemical Society}}
  \textbf{\bibinfo{volume}{138}}, \bibinfo{pages}{3410--3415}
  (\bibinfo{year}{2016}).

\bibitem{FeSC:RbEuFeAs_36K_YiLiu}
\bibinfo{author}{Liu, Y.} \emph{et~al.}
\newblock \bibinfo{title}{Superconductivity and ferromagnetism in hole-doped
  ${\mathrm{RbEuFe}}_{4}{\mathrm{As}}_{4}$}.
\newblock \emph{\bibinfo{journal}{Phys. Rev. B}} \textbf{\bibinfo{volume}{93}},
  \bibinfo{pages}{214503} (\bibinfo{year}{2016}).

\bibitem{JPSJ.85.064710}
\bibinfo{author}{Kawashima, K.} \emph{et~al.}
\newblock \bibinfo{title}{Superconductivity in fe-based compound EuAFe$_4$As$_4$ (A =
  Rb and Cs)}.
\newblock \emph{\bibinfo{journal}{Journal of the Physical Society of Japan}}
  \textbf{\bibinfo{volume}{85}}, \bibinfo{pages}{064710}
  (\bibinfo{year}{2016}).

\bibitem{PhysRevB.96.224510}
\bibinfo{author}{Liu, Y.} \emph{et~al.}
\newblock
  \bibinfo{title}{$\mathrm{RbEu}{({\mathrm{Fe}}_{1\ensuremath{-}x}{\mathrm{Ni}}_{x})}_{4}{\mathrm{As}}_{4}$:
  From a ferromagnetic superconductor to a superconducting ferromagnet}.
\newblock \emph{\bibinfo{journal}{Phys. Rev. B}} \textbf{\bibinfo{volume}{96}},
  \bibinfo{pages}{224510} (\bibinfo{year}{2017}).

\bibitem{PhysRevB.93.094513}
\bibinfo{author}{Guguchia, Z.} \emph{et~al.}
\newblock \bibinfo{title}{Probing the pairing symmetry in the over-doped
  Fe-based superconductor
  ${\mathrm{Ba}}_{0.35}{\mathrm{Rb}}_{0.65}{\mathrm{Fe}}_{2}{\mathrm{As}}_{2}$
  as a function of hydrostatic pressure}.
\newblock \emph{\bibinfo{journal}{Phys. Rev. B}} \textbf{\bibinfo{volume}{93}},
  \bibinfo{pages}{094513} (\bibinfo{year}{2016}).

\bibitem{CaKFeAs_Lochner}
\bibinfo{author}{Lochner, F.}, \bibinfo{author}{Ahn, F.},
  \bibinfo{author}{Hickel, T.} \& \bibinfo{author}{Eremin, I.}
\newblock \bibinfo{title}{Electronic properties, low-energy hamiltonian, and
  superconducting instabilities in ${\mathrm{CaKFe}}_{4}{\mathrm{As}}_{4}$}.
\newblock \emph{\bibinfo{journal}{Phys. Rev. B}} \textbf{\bibinfo{volume}{96}},
  \bibinfo{pages}{094521} (\bibinfo{year}{2017}).

\bibitem{PhysRevB.81.014526}
\bibinfo{author}{Chen, F.} \emph{et~al.}
\newblock \bibinfo{title}{Electronic structure of
  ${\text{Fe}}_{1.04}{\text{Te}}_{0.66}{\text{Se}}_{0.34}$}.
\newblock \emph{\bibinfo{journal}{Phys. Rev. B}} \textbf{\bibinfo{volume}{81}},
  \bibinfo{pages}{014526} (\bibinfo{year}{2010}).

\bibitem{PhysRevB.83.054510}
\bibinfo{author}{Zhang, Y.} \emph{et~al.}
\newblock \bibinfo{title}{Orbital characters of bands in the iron-based
  superconductor BaFe${}_{1.85}$Co${}_{0.15}$As${}_{2}$}.
\newblock \emph{\bibinfo{journal}{Phys. Rev. B}} \textbf{\bibinfo{volume}{83}},
  \bibinfo{pages}{054510} (\bibinfo{year}{2011}).

\bibitem{DFT_CaRbFeAs}
\bibinfo{author}{Shi, X.} \& \bibinfo{author}{Wang, G.}
\newblock \bibinfo{title}{Electronic structure and magnetism of the multiband
  new superconductor CarRbFe$_4$As$_4$}.
\newblock \emph{\bibinfo{journal}{Journal of the Physical Society of Japan}}
  \textbf{\bibinfo{volume}{85}}, \bibinfo{pages}{124714}
  (\bibinfo{year}{2016}).

\bibitem{PhysRevB.74.195118}
\bibinfo{author}{Wang, X.}, \bibinfo{author}{Yates, J.~R.},
  \bibinfo{author}{Souza, I.} \& \bibinfo{author}{Vanderbilt, D.}
\newblock \bibinfo{title}{Ab initio}.
\newblock \emph{\bibinfo{journal}{Phys. Rev. B}} \textbf{\bibinfo{volume}{74}},
  \bibinfo{pages}{195118} (\bibinfo{year}{2006}).

\bibitem{PhysRevB.94.064501}
\bibinfo{author}{Meier, W.~R.} \emph{et~al.}
\newblock \bibinfo{title}{Anisotropic thermodynamic and transport properties of
  single-crystalline ${\mathrm{CaKFe}}_{4}{\mathrm{As}}_{4}$}.
\newblock \emph{\bibinfo{journal}{Phys. Rev. B}} \textbf{\bibinfo{volume}{94}},
  \bibinfo{pages}{064501} (\bibinfo{year}{2016}).

\bibitem{0953-8984-23-6-065701}
\bibinfo{author}{Nowik, I.}, \bibinfo{author}{Felner, I.},
  \bibinfo{author}{Ren, Z.}, \bibinfo{author}{Cao, G.~H.} \&
  \bibinfo{author}{Xu, Z.~A.}
\newblock \bibinfo{title}{Coexistence of ferromagnetism and superconductivity:
  magnetization and mössbauer studies of EuFe$_2$(As$_{1\ensuremath{-}x}$P$_x$)$_2$}.
\newblock \emph{\bibinfo{journal}{Journal of Physics: Condensed Matter}}
  \textbf{\bibinfo{volume}{23}}, \bibinfo{pages}{065701}
  (\bibinfo{year}{2011}).

\bibitem{PhysRevB.79.155103}
\bibinfo{author}{Wu, D.} \emph{et~al.}
\newblock \bibinfo{title}{Effects of magnetic ordering on dynamical
  conductivity: Optical investigations of ${\text{EuFe}}_{2}{\text{As}}_{2}$
  single crystals}.
\newblock \emph{\bibinfo{journal}{Phys. Rev. B}} \textbf{\bibinfo{volume}{79}},
  \bibinfo{pages}{155103} (\bibinfo{year}{2009}).

\bibitem{PhysRevB.82.094426}
\bibinfo{author}{Feng, C.} \emph{et~al.}
\newblock \bibinfo{title}{Magnetic ordering and dense Kondo behavior in
  ${\text{EuFe}}_{2}{\text{P}}_{2}$}.
\newblock \emph{\bibinfo{journal}{Phys. Rev. B}} \textbf{\bibinfo{volume}{82}},
  \bibinfo{pages}{094426} (\bibinfo{year}{2010}).

\bibitem{DONIACH1977231}
\bibinfo{author}{Doniach, S.}
\newblock \bibinfo{title}{The kondo lattice and weak antiferromagnetism}.
\newblock \emph{\bibinfo{journal}{Physica B+C}} \textbf{\bibinfo{volume}{91}},
  \bibinfo{pages}{231 -- 234} (\bibinfo{year}{1977}).

\bibitem{RKKY_CeFeAsPO}
\bibinfo{author}{Luo, Y.} \emph{et~al.}
\newblock \bibinfo{title}{Phase diagram of
  ${\text{CeFeAs}}_{1\ensuremath{-}x}{\text{P}}_{x}\text{O}$ obtained from
  electrical resistivity, magnetization, and specific heat measurements}.
\newblock \emph{\bibinfo{journal}{Phys. Rev. B}} \textbf{\bibinfo{volume}{81}},
  \bibinfo{pages}{134422} (\bibinfo{year}{2010}).

\bibitem{RKKY_EuFeAsP}
\bibinfo{author}{Li, W.}, \bibinfo{author}{Zhu, J.-X.}, \bibinfo{author}{Chen,
  Y.} \& \bibinfo{author}{Ting, C.~S.}
\newblock \bibinfo{title}{First-principles calculations of the electronic
  structure of iron-pnictide EuFe${}_{2}$(As,P)${}_{2}$ superconductors:
  Evidence for antiferromagnetic spin order}.
\newblock \emph{\bibinfo{journal}{Phys. Rev. B}} \textbf{\bibinfo{volume}{86}},
  \bibinfo{pages}{155119} (\bibinfo{year}{2012}).

\bibitem{SchriefferWolff}
\bibinfo{author}{Schrieffer, J.~R.} \& \bibinfo{author}{Wolff, P.~A.}
\newblock \bibinfo{title}{Relation between the Anderson and Kondo
  hamiltonians}.
\newblock \emph{\bibinfo{journal}{Phys. Rev.}} \textbf{\bibinfo{volume}{149}},
  \bibinfo{pages}{491--492} (\bibinfo{year}{1966}).

\bibitem{PhysRevB.80.020505}
\bibinfo{author}{Dai, J.}, \bibinfo{author}{Zhu, J.-X.} \& \bibinfo{author}{Si,
  Q.}
\newblock \bibinfo{title}{$f$-spin physics of rare-earth iron pnictides:
  Influence of $d$-electron antiferromagnetic order on the heavy-fermion phase
  diagram}.
\newblock \emph{\bibinfo{journal}{Phys. Rev. B}} \textbf{\bibinfo{volume}{80}},
  \bibinfo{pages}{020505} (\bibinfo{year}{2009}).

\bibitem{PhysRevB.84.134513}
\bibinfo{author}{Akbari, A.}, \bibinfo{author}{Eremin, I.} \&
  \bibinfo{author}{Thalmeier, P.}
\newblock \bibinfo{title}{Rkky interaction in the spin-density-wave phase of
  iron-based superconductors}.
\newblock \emph{\bibinfo{journal}{Phys. Rev. B}} \textbf{\bibinfo{volume}{84}},
  \bibinfo{pages}{134513} (\bibinfo{year}{2011}).

\bibitem{EuFeAs_ZhiRen}
\bibinfo{author}{Ren, Z.} \emph{et~al.}
\newblock \bibinfo{title}{Suppression of spin-density-wave transition and
  emergence of ferromagnetic ordering of ${\text{Eu}}^{2+}$ moments in
  ${\text{EuFe}}_{2\ensuremath{-}x}{\text{Ni}}_{x}{\text{As}}_{2}$}.
\newblock \emph{\bibinfo{journal}{Phys. Rev. B}} \textbf{\bibinfo{volume}{79}},
  \bibinfo{pages}{094426} (\bibinfo{year}{2009}).

\bibitem{PhysRevLett.113.087202}
\bibinfo{author}{Yao, N.~Y.}, \bibinfo{author}{Glazman, L.~I.},
  \bibinfo{author}{Demler, E.~A.}, \bibinfo{author}{Lukin, M.~D.} \&
  \bibinfo{author}{Sau, J.~D.}
\newblock \bibinfo{title}{Enhanced antiferromagnetic exchange between magnetic
  impurities in a superconducting host}.
\newblock \emph{\bibinfo{journal}{Phys. Rev. Lett.}}
  \textbf{\bibinfo{volume}{113}}, \bibinfo{pages}{087202}
  (\bibinfo{year}{2014}).

\bibitem{PhysRevLett.101.057003}
\bibinfo{author}{Mazin, I.~I.}, \bibinfo{author}{Singh, D.~J.},
  \bibinfo{author}{Johannes, M.~D.} \& \bibinfo{author}{Du, M.~H.}
\newblock \bibinfo{title}{Unconventional superconductivity with a sign reversal
  in the order parameter of
  ${\mathrm{LaFeAsO}}_{1\ensuremath{-}x}{\mathrm{F}}_{x}$}.
\newblock \emph{\bibinfo{journal}{Phys. Rev. Lett.}}
  \textbf{\bibinfo{volume}{101}}, \bibinfo{pages}{057003}
  (\bibinfo{year}{2008}).

\bibitem{doi:10.1143/JPSJ.77.114709}
\bibinfo{author}{Kitagawa, K.}, \bibinfo{author}{Katayama, N.},
  \bibinfo{author}{Ohgushi, K.}, \bibinfo{author}{Yoshida, M.} \&
  \bibinfo{author}{Takigawa, M.}
\newblock \bibinfo{title}{Commensurate itinerant antiferromagnetism in
  BaFe$_2$As$_2$: 75As-NMR studies on a self-flux grown single crystal}.
\newblock \emph{\bibinfo{journal}{Journal of the Physical Society of Japan}}
  \textbf{\bibinfo{volume}{77}}, \bibinfo{pages}{114709}
  (\bibinfo{year}{2008}).

\bibitem{PhysRevB.82.180501}
\bibinfo{author}{Ma, L.}, \bibinfo{author}{Zhang, J.}, \bibinfo{author}{Chen,
  G.~F.} \& \bibinfo{author}{Yu, W.}
\newblock \bibinfo{title}{NMR evidence of strongly correlated superconductivity
  in LiFeAs: Tuning toward a spin-density-wave ordering}.
\newblock \emph{\bibinfo{journal}{Phys. Rev. B}} \textbf{\bibinfo{volume}{82}},
  \bibinfo{pages}{180501} (\bibinfo{year}{2010}).

\bibitem{QE}
\bibinfo{author}{Giannozzi, P.} \emph{et~al.}
\newblock \bibinfo{title}{Quantum espresso: a modular and open-source software
  project for quantum simulations of materials}.
\newblock \emph{\bibinfo{journal}{Journal of Physics: Condensed Matter}}
  \textbf{\bibinfo{volume}{21}}, \bibinfo{pages}{395502}
  (\bibinfo{year}{2009}).

\bibitem{PhysRevB.54.11169}
\bibinfo{author}{Kresse, G.} \& \bibinfo{author}{Furthm\"uller, J.}
\newblock \bibinfo{title}{Efficient iterative schemes for ab initio
  total-energy calculations using a plane-wave basis set}.
\newblock \emph{\bibinfo{journal}{Phys. Rev. B}} \textbf{\bibinfo{volume}{54}},
  \bibinfo{pages}{11169--11186} (\bibinfo{year}{1996}).

\bibitem{PhysRevB.59.1758}
\bibinfo{author}{Kresse, G.} \& \bibinfo{author}{Joubert, D.}
\newblock \bibinfo{title}{From ultrasoft pseudopotentials to the projector
  augmented-wave method}.
\newblock \emph{\bibinfo{journal}{Phys. Rev. B}} \textbf{\bibinfo{volume}{59}},
  \bibinfo{pages}{1758--1775} (\bibinfo{year}{1999}).

\bibitem{GGA}
\bibinfo{author}{Perdew, J.~P.}, \bibinfo{author}{Burke, K.} \&
  \bibinfo{author}{Ernzerhof, M.}
\newblock \bibinfo{title}{Generalized gradient approximation made simple}.
\newblock \emph{\bibinfo{journal}{Phys. Rev. Lett.}}
  \textbf{\bibinfo{volume}{77}}, \bibinfo{pages}{3865--3868}
  (\bibinfo{year}{1996}).

\bibitem{Monkhorst-Pack}
\bibinfo{author}{Monkhorst, H.~J.} \& \bibinfo{author}{Pack, J.~D.}
\newblock \bibinfo{title}{Special points for Brillouin-zone integrations}.
\newblock \emph{\bibinfo{journal}{Phys. Rev. B}} \textbf{\bibinfo{volume}{13}},
  \bibinfo{pages}{5188--5192} (\bibinfo{year}{1976}).

\bibitem{LDAU}
\bibinfo{author}{Anisimov, V.~I.}, \bibinfo{author}{Zaanen, J.} \&
  \bibinfo{author}{Andersen, O.~K.}
\newblock \bibinfo{title}{Band theory and Mott insulators: Hubbard $U$ instead of
  stoner $I$}.
\newblock \emph{\bibinfo{journal}{Phys. Rev. B}} \textbf{\bibinfo{volume}{44}},
  \bibinfo{pages}{943--954} (\bibinfo{year}{1991}).

\bibitem{DFTU}
\bibinfo{author}{Anisimov, V.~I.} \& \bibinfo{author}{Gunnarsson, O.}
\newblock \bibinfo{title}{Density-functional calculation of effective Coulomb
  interactions in metals}.
\newblock \emph{\bibinfo{journal}{Phys. Rev. B}} \textbf{\bibinfo{volume}{43}},
  \bibinfo{pages}{7570--7574} (\bibinfo{year}{1991}).

\bibitem{LSDAU}
\bibinfo{author}{Antonov, V.~N.}, \bibinfo{author}{Harmon, B.~N.} \&
  \bibinfo{author}{Yaresko, A.~N.}
\newblock \bibinfo{title}{Electronic structure of mixed-valence and
  charge-ordered Sm and Eu pnictides and chalcogenides}.
\newblock \emph{\bibinfo{journal}{Phys. Rev. B}} \textbf{\bibinfo{volume}{72}},
  \bibinfo{pages}{085119} (\bibinfo{year}{2005}).

\bibitem{PhysRevLett.100.237003}
\bibinfo{author}{Singh, D.~J.} \& \bibinfo{author}{Du, M.-H.}
\newblock \bibinfo{title}{Density functional study of
  ${\mathrm{LaFeAsO}}_{1\ensuremath{-}x}{\mathrm{F}}_{x}$: A low carrier
  density superconductor near itinerant magnetism}.
\newblock \emph{\bibinfo{journal}{Phys. Rev. Lett.}}
  \textbf{\bibinfo{volume}{100}}, \bibinfo{pages}{237003}
  (\bibinfo{year}{2008}).

\bibitem{Ma2010}
\bibinfo{author}{Ma, F.-j.}, \bibinfo{author}{Lu, Z.-Y.} \&
  \bibinfo{author}{Xiang, T.}
\newblock \bibinfo{title}{Electronic structures of ternary iron arsenides
  AFe$_2$As$_2$ (A = Ba, Ca, or Sr)}.
\newblock \emph{\bibinfo{journal}{Frontiers of Physics in China}}
  \textbf{\bibinfo{volume}{5}}, \bibinfo{pages}{150--160}
  (\bibinfo{year}{2010}).

\bibitem{PhysRevB.78.085104}
\bibinfo{author}{Mazin, I.~I.}, \bibinfo{author}{Johannes, M.~D.},
  \bibinfo{author}{Boeri, L.}, \bibinfo{author}{Koepernik, K.} \&
  \bibinfo{author}{Singh, D.~J.}
\newblock \bibinfo{title}{Problems with reconciling density functional theory
  calculations with experiment in ferropnictides}.
\newblock \emph{\bibinfo{journal}{Phys. Rev. B}} \textbf{\bibinfo{volume}{78}},
  \bibinfo{pages}{085104} (\bibinfo{year}{2008}).

\bibitem{Wannier90_code}
\bibinfo{author}{Mostofi, A.~A.} \emph{et~al.}
\newblock \bibinfo{title}{wannier90: A tool for obtaining maximally-localised
  wannier functions}.
\newblock \emph{\bibinfo{journal}{Computer Physics Communications}}
  \textbf{\bibinfo{volume}{178}}, \bibinfo{pages}{685 -- 699}
  (\bibinfo{year}{2008}).

\end{thebibliography}

\newpage

\newpage 

\begin{table}
\begin{tabular}{c||c|c|cc}
\hline
 & BaFe$_2$As$_2$ &  CaRbFe$_{4}$As$_{4}$ &\multicolumn{2}{c }{EuRbFe$_{4}$As$_{4}$} \\
 & & & PM & FM  \\
\hline
$3d_{z^{2}}-4p_z$ & 0.544  & 0.478 & 0.544 & 0.545 \\
$3d_{z^{2}}-4p_x$ & 0.186  & 0.247 & 0.220 & 0.221 \\
$3d_{xz}-4p_z$    & 0.029  & 0.106 & 0.059 & 0.061 \\
$3d_{xz}-4p_x$    & 0.179  & 0.143 & 0.170 & 0.171 \\
$3d_{xz}-4p_y$    & 0.759  & 0.689 & 0.747 & 0.751 \\
$3d_{x^2}-4p_x$     & 0.556 & 0.490 & 0.541 & 0.543 \\
$3d_{xy}-4p_z$     & 0.678 & 0.587 & 0.623 & 0.624 \\
$3d_{xy}-4p_x$     & 0.292 & 0.168 & 0.216 & 0.217 \\
\hline
$3d_{z^{2}}-4p'_z$ & 0.544 & 0.541 & 0.515 & 0.521 \\
$3d_{z^{2}}-4p'_x$ & 0.186 & 0.260 & 0.244 & 0.247 \\
$3d_{xz}-4p'_z$    & 0.029 & 0.096 & 0.092 & 0.093 \\
$3d_{xz}-4p'_x$    & 0.179 & 0.155 & 0.156 & 0.153 \\
$3d_{xz}-4p'_y$    & 0.759 & 0.748 & 0.755 & 0.762 \\
$3d_{x^2}-4p'_x$   & 0.556 & 0.514 & 0.526 & 0.526 \\
$3d_{xy}-4p'_z$     & 0.678 & 0.640 & 0.658 & 0.667 \\
$3d_{xy}-4p'_x$     & 0.292 & 0.162 & 0.206 & 0.205 \\
\hline
$4p_z-4p'_z$        & 0.335 & 0.360 & 0.349 & 0.347 \\
$4p_x-4p'_x$        & 0.570 & 0.362 & 0.440 & 0.448 \\
$4p_y-4p'_y$        & 0.147 & 0.197 & 0.196 & 0.209 \\
\hline
$3d_{z^{2}}-3d'_{z^{2}}$ & 0.108 & 0.137 & 0.132 & 0.134 \\
$3d_{xz}-3d'_{xz}$ & 0.015 & 0.028 & 0.022 & 0.022 \\
$3d_{yz}-3d'_{yz}$ & 0.214 & 0.280 & 0.265 & 0.266 \\
$3d_{x^2}-3d'_{x^2}$ &  0.346 & 0.363& 0.367& 0.366 \\
$3d_{xy}-3d'_{xy}$  &  0.182 & 0.276 & 0.254 & 0.254 \\
$3d_{z^{2}}-f_{z^{3}}$ & - & - & - & 0.033  \\
$3d_{z^{2}}-f_{xz^{2}}$& - & - & - & 0.024  \\
$3d_{z^{2}}-f_{xyz}$& - & - & - & 0.051  \\
$3d_{z^{2}}-f_{3xy^{2}}$& - & - & - & 0.015  \\
$3d_{xz}-f_{z^{3}}$& - & - & - & 0.032  \\
$3d_{xz}-f_{xz^{2}}$& - & - & - & 0.015  \\
$3d_{xz}-f_{yz^{2}}$& - & - & - & 0.037  \\
$3d_{x^2}-f_{xz^{2}}$& - & - & - & 0.029  \\
\hline
\end{tabular}
\caption{Nearest neighboring hoppings $t_{ij}$ (in eV) from maximally projected Wannier function method (only symmetrically inequivalent terms larger than 1 meV are shown). In 1144-type, the hoppings between Fe-3$d$ and As-4$p$ will be different for As$^{\mathrm{I}}$ (denoted with $p$) and As$^{\mathrm{II}}$ (denoted with $p'$) (Fig. \ref{fig:CS_CEF}b) due to the local S$_4$ symmetry broken. \label{tab:hopping}}
\end{table}

\begin{table}
\begin{tabular}{c|cccc}
\hline
     &  BaFe$_2$As$_2$ $\quad$ & $\quad$ CaRbFe$_4$As$_4$  $\quad$
     & $\quad$ EuRbFe$_4$As$_4$  & $\quad$ EuRbNi$_4$As$_4$ \\
\hline
 $\quad d_{z^{2}} $       & -0.730  & -0.780  & -0.730 & -1.948  \\

 $\quad d_{xz} / d_{yz} $ &  -0.659  & -0.730  & -0.678 & -1.908 \\

 $\quad d_{x^{2}-y^{2}} $ & -0.914  & -0.985 &  -0.962 & -2.061\\

 $\quad d_{xy}          $ & -0.700  & -0.774  & -0.746 & -1.906\\

 $\Delta_{cf}$ & 0.128  &  0.127  &  0.142  & 0.078 \\

 $\Delta_{ex}$ & 0.00  &  0.00  &  0.012   & 0.020\\
\hline
\end{tabular}
\caption{The crystal field splitting and exchange splitting (in eV) of Fe-$3d$ orbitals with respect to Fermi level for BaFe$_2$As$_2$, CaRbFe$_4$As$_4$, EuRbFe$_4$As$_4$ and EuRbNi$_4$As$_4$ (100\% Ni doping).\label{tab:splitting}}
\end{table}

\begin{table}[h]
\begin{tabular}{c||c|c|c|c}
\hline    
 $x$         &  0\%   & 6.25\% & 12.5\%   & 100\%  \\
 \hline
$\bar{J_K}$ (meV)& 0.730  & 0.605(0.604)       & 0.662(0.510)         & 0.570   \\
$J_{\mathrm{RKKY}} (meV)$        & 0.120  & 0.093(0.117)       & 0.100(0.086)         & 0.070   \\
\hline
\end{tabular}
\caption{The variation of averaged Kondo coupling $\bar{J_K}$, and RKKY interaction $J_{\mathrm{RKKY}}$ in EuRb(Fe$_{1-x}$N$_{x}$As)$_4$ with respect to doping level $x$. Note that in experiment the Ni-doping is within 10\% \cite{PhysRevB.96.224510}. For doped system ($0<x<1$), the numbers inside/outside the bracket were calculated using RBM/VCA, respectively.\label{tab:RKKY}}
\end{table}

\end{document}